\documentclass[pra,aps,amsmath,amssymb,showpacs,showkeys]{revtex4}
\usepackage{graphicx}

\newcommand{\bsg}{\boldsymbol{\sigma}}
\newcommand{\bro}{\boldsymbol{\rho}}

\newcommand{\pen}{\openone}

\newcommand{\wth}{\widetilde{H}}
\newcommand{\wff}{\widetilde{F}}

\begin{document}
\clearpage
\preprint{}

\title{Uncertainty and certainty relations for complementary qubit observables in terms of Tsallis' entropies}
\author{Alexey E. Rastegin}
\affiliation{Department of Theoretical Physics, Irkutsk State University,
Gagarin Bv. 20, Irkutsk 664003, Russia}

\begin{abstract}
Uncertainty relations for more than two observables have found use
in quantum information, though commonly known relations pertain to
a pair of observables. We present novel uncertainty and certainty
relations of state-independent form for the three Pauli
observables with use of the Tsallis $\alpha$-entropies. For all
real $\alpha\in(0;1]$ and integer $\alpha\geq2$, lower bounds on
the sum of three $\alpha$-entropies are obtained. These bounds are
tight in the sense that they are always reached with certain pure
states. The necessary and sufficient condition for equality is
that the qubit state is an eigenstate of one of the Pauli
observables. Using concavity with respect to the parameter
$\alpha$, we derive approximate lower bounds for non-integer
$\alpha\in(1;+\infty)$. In the case of pure states, the developed
method also allows to obtain upper bounds on the entropic sum for
real $\alpha\in(0;1]$ and integer $\alpha\geq2$. For applied
purposes, entropic bounds are often used with averaging over the
individual entropies. Combining the obtained bounds leads to a
band, in which the rescaled average $\alpha$-entropy ranges in the
pure-state case. A width of this band is essentially dependent on
$\alpha$. It can be interpreted as an evidence for sensitivity in
quantifying the complementarity.
\end{abstract}


\maketitle

\section{Introduction}\label{sec1}

The quantum-mechanical concept of complementarity is naturally
posed by means of uncertainty relations. Since original
Heisenberg's result \cite{heisenberg} had been given, numerous
ways to pose the uncertainty principle were proposed
\cite{hall99,lahti}. Historically, uncertainty relations were
focused on pairs of canonically conjugate variables. Today,
uncertainty relations attract an attention also due to potential
applications in quantum information processing
\cite{dmg07,renes09,vsw09}. Entropic functions provide a flexible
tool for expressing an uncertainty in quantum measurements. In
view of the existing reviews \cite{ww10,brud11}, we mention here
only selected developments. The most traditional form of
uncertainty relations was proposed by Robertson \cite{robert}.
Deutsch has discussed advantages of the entropic approach
\cite{deutsch}. For the case of two measurements, the inequality
of Maassen and Uffink \cite{maass} is widely used. This advance
has been inspired by the previous conjecture of Kraus
\cite{kraus87}. It is based on a deep mathematical result known as
Riesz's theorem. In this way, the Maassen--Uffink relation can be
extended to a pair of POVM measurements \cite{krpr02,rast102}.

Entropic inequalities of the Maassen--Uffink type have emphasized
a role of mutual unbiasedness. In this regard, we can ask for the
entropic uncertainty tradeoff between more than two observables
\cite{ww10,aamb10}. Entropic uncertainty bounds for several
observables are also of interest in studying the security of
quantum cryptographic protocols \cite{dmg07,ngbw12}. Uncertainties
are mainly quantified by means of the Shannon entropy. An entropic
uncertainty relation for $(d+1)$ mutually unbiased bases in
$d$-dimensional Hilbert space was obtained \cite{ivan92,jsan93}
and improved \cite{jsan95}. The case of arbitrary number of
mutually unbiased bases was examined in \cite{molm09}. The writer
of \cite{jsan93} also gave the exact bounds for the qubit case
$d=2$. In \cite{ww08}, uncertainty relations for a set of
anti-commuting observables were given in terms of the Shannon
entropy and the so-called collision one (R\'{e}nyi's entropy of
order 2). Although the Shannon entropy is of great importance,
both the R\'{e}nyi \cite{renyi61} and Tsallis \cite{tsallis}
entropies have found use in many issues.

For the traditional case of canonically conjugate observables, the
formulation in terms of R\'{e}nyi's entropies was given in
\cite{birula3}. Both the R\'{e}nyi and Tsallis entropies have been
used in expressing the uncertainty principle for trace-preserving
super-operators \cite{rast104} and the number and annihilation
operators \cite{rast105}. Reformulations of the entropic
uncertainty principle in the presence of quantum memory are
considered in \cite{BCCRR10,clz11,fan12}. For quasi-Hermitian
models, uncertainty relations have been derived in terms of the
so-called unified entropies \cite{rastnnh}. The R\'{e}nyi and
Tsallis entropies are both included in the family of unified
entropies proposed in \cite{hey06}. In the papers
\cite{lars90,diaz}, the complementarity issue is examined with the
sum of squares of probabilities. This sum is closely related to
Tsallis' $2$-entropy. Basing on the Riesz theorem, unified-entropy
uncertainty relations for various pairs of measurement have been
obtained in \cite{rast12qic}. In \cite{rastijtp}, we considered
entropic inequalities beyond the scope of Riesz's theorem. For
some pairs of measurements, uncertainty relations were posed in
terms of both the R\'{e}nyi and Tsalis $1/2$-entropies
\cite{rastijtp}.

In the present work, we consider lower and upper bounds on the sum
of Tsallis' $\alpha$-entropies, which quantify uncertainties in
measurement of complementary qubit observables. These observables
are commonly represented by the Pauli matrices. Results of such a
kind may be useful in studying the security of six-state protocols
of quantum key distribution. From this viewpoint, some uncertainty
relations in information-theoretical terms are extensively treated
in \cite{dmg07,ngbw12,bfw11}. The paper is organized as follows.
The preliminary material is given in Sect. \ref{sec2}. In Sect.
\ref{sec3}, tight lower bounds on the sum of three entropies of
degree $\alpha\in(0;1]$ are obtained. The conditions for equality
are considered as well. Tight lower bounds on the entropic sum for
integer $\alpha\geq2$ are derived in Sect. \ref{sec4}. Using
concavity with respect to the parameter $\alpha$, we also obtain
approximate lower bounds for non-integer $\alpha\in(1;+\infty)$.
In Sect. \ref{sec5}, we examine upper bounds on the sum of three
$\alpha$-entropies in the case of pure states. Here, the bounds
are given for all real $\alpha\in(0;1]$ and integer
$\alpha\geq2$. In Sect. \ref{sec6}, we conclude the paper with a
summary of results.

\section{Definition and notation}\label{sec2}

In this section, the preliminary material is given. In the
considered approach, uncertainties of quantum measurements are
quantified by means of entropies. In the following, we use the
Tsallis entropy. Let $p=\{p_{j}\}$ be a probability distribution
supported on $n$ points. For real $\alpha>0\neq1$, the
non-extensive $\alpha$-entropy is defined by \cite{tsallis}
\begin{equation}
H_{\alpha}(p):=\frac{1}{1-\alpha}{\,}\left(\sum\nolimits_{j=1}^{n} p_{j}^{\alpha}-1\right)=
\sum\nolimits_{j=1}^{n} h_{\alpha}(p_{j})
\ . \label{tslent}
\end{equation}
For brevity, we introduce here the function
\begin{equation}
h_{\alpha}(u):=\frac{u^{\alpha}-u}{1-\alpha}=-u^{\alpha}\ln_{\alpha}(u)
\ , \label{ahdf}
\end{equation}
and the $\alpha$-logarithm
$\ln_{\alpha}(u):=\bigl(u^{1-\alpha}-1\bigr)/(1-\alpha)$. When
$\alpha\to1$, we obtain the usual logarithm $\ln{u}$ and the
Shannon entropy
\begin{equation}
H_{1}(p)=\sum\nolimits_{j=1}^{n}h_{1}(p_{j})=-\sum\nolimits_{j=1}^{n}p_{j}\ln{p}_{j}
\ . \label{shent}
\end{equation}
With the factor $\left(2^{1-\alpha}-1\right)^{-1}$ instead of
$(1-\alpha)^{-1}$, the entropic function (\ref{tslent}) was
derived from several axioms by Havrda and Charv\'{a}t \cite{HC67}.
In the context of statistical physics, the entropy (\ref{tslent})
was developed by Tsallis \cite{tsallis}.

The $\alpha$-entropy (\ref{tslent}) is a concave function of
probability distribution. Namely, for all $\lambda\in[0;1]$ and
two probability distributions $p=\{p_{j}\}$ and $q=\{q_{j}\}$, we
have
\begin{equation}
H_{\alpha}\bigl(\lambda{p}+(1-\lambda)q\bigr)\geq
\lambda{\,}H_{\alpha}(p)+(1-\lambda)H_{\alpha}(q)
\ . \label{hacn}
\end{equation}
This follows from concavity of the function (\ref{ahdf}) with
respect to the variable $u$. Combining this property with
$h_{\alpha}(0)=h_{\alpha}(1)=0$, we also get $h_{\alpha}(u)\geq0$.
Tsallis' $\alpha$-entropy (\ref{tslent}) satisfies
\begin{equation}
0\leq{H}_{\alpha}(p)\leq\ln_{\alpha}(n)
\ . \label{hamm}
\end{equation}
The maximum $\ln_{\alpha}(n)$ is reached with the equiprobable
distribution, i.e. $p_{j}=1/n$ for all $1\leq{j}\leq{n}$. The
minimal zero value is reached with any deterministic distribution,
when one of probabilities is 1 and other are all zeros. In fact,
the values $u=0$ and $u=1$ are only ones, for which the function
$h_{\alpha}(u)$ vanishes. This is actually a manifestation of the
fact that the function is strictly concave. Indeed, for $u>0$ we
have the relation
$h_{\alpha}^{\prime\prime}(u)=-\alpha{u}^{\alpha-2}<0$ sufficient
for strict concavity. Following the paper \cite{rastijtp}, we also
put the quantity
\begin{equation}
\Phi_{\alpha}(p):=\sum\nolimits_{j}p_j^{\alpha}
\ . \label{prerm}
\end{equation}
As $p_{j}\leq1$, we have $\Phi_{\alpha}(p)\geq\Phi_{\beta}(p)$
whenever $\alpha<\beta$. With respect to the distribution
$p=\{p_{j}\}$, the functional (\ref{prerm}) is concave for
$\alpha\in(0;1)$ and convex for $\alpha\in(1;+\infty)$. Using this
functional, we represent the entropy (\ref{tslent}) as
\begin{equation}
H_{\alpha}(p)=\frac{\Phi_{\alpha}(p)-1}{1-\alpha}
\ . \label{taph}
\end{equation}
R\'{e}nyi entropies form another especially important family of
one-parametric extensions of the Shannon entropy. Basic properties
of this extension are examined in the original work
\cite{renyi61}. Some of applications of the R\'{e}nyi and Tsallis
entropies in quantum information theory are discussed in
\cite{bengtsson}.

In the present paper, we will deal with the qubit case $n=2$. In
this case, three complementary observables are usually represented
by the Pauli matrices $\bsg_{x}$, $\bsg_{y}$, $\bsg_{z}$, namely
\begin{equation}
\bsg_{x}=
\begin{pmatrix}
0 & 1 \\
1 & 0
\end{pmatrix}
\ , \qquad
\bsg_{y}=
\begin{pmatrix}
0 & -i \\
i & 0
\end{pmatrix}
\ , \qquad
\bsg_{z}=
\begin{pmatrix}
1 & 0 \\
0 & -1
\end{pmatrix}
\ . \label{pmtr}
\end{equation}
Historically, these matrices were introduced in describing
spin-$1/2$ observables. Each of the matrices has the eigenvalues
$\pm1$. Let $\bigl\{|0\rangle,|1\rangle\bigr\}$ denote the
eigenbasis of $\bsg_{z}$, in the matrix form
\begin{equation}
|0\rangle=
\begin{pmatrix}
1  \\
0
\end{pmatrix}
\ , \qquad
|1\rangle=
\begin{pmatrix}
0  \\
1
\end{pmatrix}
\ . \label{01bs}
\end{equation}
The normalized eigenvectors of $\bsg_{x}$ and $\bsg_{y}$ can be
written as
\begin{equation}
|x_{\pm}\rangle=\frac{1}{\sqrt{2}}
\begin{pmatrix}
1  \\
\pm1
\end{pmatrix}
\ , \qquad
|y_{\pm}\rangle=\frac{1}{\sqrt{2}}
\begin{pmatrix}
1  \\
\pm{i}
\end{pmatrix}
\ . \label{xybb}
\end{equation}
We obviously have $\bsg_{x}|x_{\pm}\rangle=\pm|x_{\pm}\rangle$ and
$\bsg_{y}|y_{\pm}\rangle=\pm|y_{\pm}\rangle$. Note that the kets
$|y_{+}\rangle$ and $|y_{-}\rangle$ correspond to the states of
right and left circular polarizations, respectively. The three
bases given by (\ref{01bs}) and (\ref{xybb}) are mutually
unbiased. Measurements in these eigenbases are used in six-state
cryptographic protocols \cite{ngbw12,bfw11}.

Let us write the probabilities corresponding to measurement of
each of the observables $\bsg_{x}$, $\bsg_{y}$, $\bsg_{z}$. Up to
a unimodular factor, we can express a normalized pure state as
\begin{equation}
|\psi\rangle=\cos\tau|0\rangle+e^{i\varphi}\sin\tau|1\rangle=
\begin{pmatrix}
\cos\tau  \\
e^{i\varphi}\sin\tau
\end{pmatrix}
\ , \label{pswr}
\end{equation}
where $\tau$ and $\varphi$ are real numbers. Assuming
$\varphi\in[0;2\pi)$, we will take $\tau\in[0;\pi/2]$, since the
reversed sign in the state vector has no physical relevance. The
probabilities are calculated as follows. With respect to the basis
$\bigl\{|x_{+}\rangle,|x_{-}\rangle\bigr\}$, we obtain
\begin{equation}
p_{\pm}=\bigl|\langle{x}_{\pm}|\psi\rangle\bigr|^{2}=\frac{1\pm\sin2\tau\cos\varphi}{2}
\ . \label{xpr}
\end{equation}
With respect to the basis
$\bigl\{|y_{+}\rangle,|y_{-}\rangle\bigr\}$, we further have
\begin{equation}
q_{\pm}=\bigl|\langle{y}_{\pm}|\psi\rangle\bigr|^{2}=\frac{1\pm\sin2\tau\sin\varphi}{2}
\ . \label{ypr}
\end{equation}
With respect to the basis $\bigl\{|0\rangle,|1\rangle\bigr\}$, we
obtain $r_{+}=\bigl|\langle0|\psi\rangle\bigr|^{2}=\cos^{2}\tau$
and $r_{-}=\bigl|\langle1|\psi\rangle\bigr|^{2}=\sin^{2}\tau$, or
merely
\begin{equation}
r_{\pm}=\frac{1\pm\cos2\tau}{2}
\ . \label{zpr}
\end{equation}
Substituting the post-measurements distributions (\ref{xpr}),
(\ref{ypr}), (\ref{zpr}) into the right-hand side of
(\ref{tslent}), we respectively obtain the entropies
$H_{\alpha}\bigl(\bsg_{x}|\psi\bigr)$,
$H_{\alpha}\bigl(\bsg_{y}|\psi\bigr)$,
$H_{\alpha}\bigl(\bsg_{z}|\psi\bigr)$ for the state (\ref{pswr}).
We will derive lower and upper bounds on the sum of such entropies
for pure states and, by concavity, for mixed ones.

\section{Tight lower bounds on the sum of entropies of degree $\alpha\in(0;1]$}\label{sec3}

In this section, we derive tight lower bounds on the entropic sum
for $\alpha\in(0;1]$. A desired bound will firstly be obtained for
pure states of the form (\ref{pswr}), when the probabilities are
given by (\ref{xpr}), (\ref{ypr}) and (\ref{zpr}). Using the
concavity properties, we then extend the result to mixed states of
a qubit. For $\alpha>0\neq1$, we introduce the function
\begin{equation}
F_{\alpha}(\tau,\varphi)=\frac{1}{1-\alpha}{\,}
\left(\sum\nolimits_{j=\pm}\left(p_{j}^{\alpha}+q_{j}^{\alpha}+r_{j}^{\alpha}\right)-3\right)
\ . \label{ftpd}
\end{equation}
This function represents the entropic sum
$H_{\alpha}\bigl(\bsg_{x}|\psi\bigr)+H_{\alpha}\bigl(\bsg_{y}|\psi\bigr)+H_{\alpha}\bigl(\bsg_{z}|\psi\bigr)$
in terms of the variables $\tau$ and $\varphi$. Formally, we aim
to minimize (\ref{ftpd}) in the domain of interest. As was noted
above, the variables are initially in the intervals
$\tau\in[0;\pi/2]$ and $\varphi\in[0;2\pi)$. In the task of
optimization, however, we can restrict a consideration to the
rectangular domain
\begin{equation}
D:=\bigl\{
(\tau,\varphi):{\>}\tau\in[0;\pi/4],{\>}\varphi\in[0;\pi/4]
\bigr\}
\ . \label{dotv}
\end{equation}
In the total domain
$\bigl\{(\tau,\varphi):{\>}\tau\in[0;\pi/2],{\>}\varphi\in[0;2\pi)\bigr\}$,
the function $F_{\alpha}(\tau,\varphi)$ takes the same range of
values as on the domain (\ref{dotv}). The justification is as
follows. First, mapping $\varphi\mapsto\varphi-\pi$ merely swaps
two values in the pairs (\ref{xpr}) and (\ref{ypr}). For each of
the pair $p_{\pm}$ and $q_{\pm}$, therefore, the interval
$\varphi\in[\pi;2\pi]$ leads to the same values as the interval
$\varphi\in[0;\pi]$. Further, under mapping
$\varphi\mapsto\pi-\varphi$ the probabilities $p_{\pm}$ are
swapped and the probabilities $q_{\pm}$ are unchanged. Hence, we
can restrict a consideration to values of
$F_{\alpha}(\tau,\varphi)$ for $\varphi\in[0;\pi/2]$. Acting in
this interval, mapping $\varphi\mapsto\pi/2-\varphi$ implies
swapping $p_{j}$ and $q_{j}$, where $j=\pm$. So, the interval
$\varphi\in[\pi/4;\pi/2]$ does not give new values of the function
$F_{\alpha}(\tau,\varphi)$, and we shall now assume
$\varphi\in[0;\pi/4]$. Finally, we use mapping
$\tau\mapsto\pi/2-\tau$, which does not alter $\sin2\tau$ and
reverses the sign of $\cos2\tau$. Hence, only the probabilities
$r_{\pm}$ are merely swapped. The following statement takes place.

\newtheorem{thm31}{Theorem}
\begin{thm31}\label{inta01}
Let qubit state be described by density matrix $\bro$. For all
$\alpha\in(0;1]$, the entropic sum satisfies
\begin{equation}
H_{\alpha}\bigl(\bsg_{x}|\bro\bigr)+H_{\alpha}\bigl(\bsg_{y}|\bro\bigr)+
H_{\alpha}\bigl(\bsg_{z}|\bro\bigr)\geq2\ln_{\alpha}(2)
\ , \label{hal0}
\end{equation}
with equality if and only if the qubit state is an eigenstate of
either of the observables $\bsg_{x}$, $\bsg_{y}$, $\bsg_{z}$.
\end{thm31}

{\bf{Proof.}} We first assume that $\alpha\neq1$. Let us show that
the right-hand side of (\ref{hal0}) gives the minimum of
(\ref{ftpd}) in the domain (\ref{dotv}). Differentiating with
respect to $\varphi$, we obtain
\begin{align}
&\frac{\partial}{\partial\varphi}{\>}F_{\alpha}(\tau,\varphi)=
-\frac{\alpha}{2(1-\alpha)}\left(p_{+}^{\alpha-1}-p_{-}^{\alpha-1}\right)\sin2\tau\sin\varphi
\nonumber\\
&+\frac{\alpha}{2(1-\alpha)}\left(q_{+}^{\alpha-1}-q_{-}^{\alpha-1}\right)\sin2\tau\cos\varphi
=\alpha2^{-\alpha}uv\bigl[f_{\alpha}(u)-f_{\alpha}(v)\bigr]
{\>}. \label{drph}
\end{align}
For brevity, we introduce here the variables
$u=\sin2\tau\cos\varphi$, $v=\sin2\tau\sin\varphi$, and the
function
\begin{equation}
f_{\alpha}(u)=\frac{(1-u)^{\alpha-1}-(1+u)^{\alpha-1}}{(1-\alpha)u}
\ . \label{gadf}
\end{equation}
Except for the boundary lines of the rectangle (\ref{dotv}), we
have $0<v<u<1$. We now claim that the function $f_{\alpha}(u)$
monotonically increases with $u\in(0;1)$ for all real
$\alpha\in(0;1)$. This fact easily follows from its expansion as a
power series about the origin. Using the binomial theorem, we
actually obtain
\begin{equation}
f_{\alpha}(u)=\sum_{k=0}^{\infty}\frac{2}{1-\alpha}{\,}\binom{2k+1-\alpha}{2k+1} u^{2k}
\ . \label{al01}
\end{equation}
We stress that this series contains only strictly positive
coefficients. In fact, for $k\geq0$ and $\alpha\in(0;1)$ we have
\begin{equation}
\frac{2}{1-\alpha}{\,}
\binom{2k+1-\alpha}{2k+1}=2{\>}\frac{(2k+1-\alpha)(2k-\alpha)\cdots(2-\alpha)}{(2k+1)!}>0
\ . \label{bi01}
\end{equation}
So the function (\ref{al01}) monotonically increases with
$u\in(0;1)$. Hence, the inequality $v<u$ implies
$f_{\alpha}(v)<f_{\alpha}(u)$. In the interior of the domain
(\ref{dotv}), therefore, the derivative (\ref{drph}) is strictly
positive. Here, the function $F_{\alpha}(\tau,\varphi)$ increases
with $\varphi$. On the boundary lines $\tau=0$ and $\tau=\pi/4$,
we have $\partial{F}_{\alpha}/\partial\varphi=0$. These facts
implies that the minimal and maximal values of $F_{\alpha}(\tau,\varphi)$
in the domain (\ref{dotv}) are reached on the lines $\varphi=0$
and $\varphi=\pi/4$, respectively. To find the minimum, we
substitute $\varphi=0$ and rewrite probabilities as
\begin{equation}
p_{\pm}=\frac{1\pm\sin2\tau}{2}
\ , \qquad
q_{\pm}=\frac{1}{2}
\ , \qquad
r_{\pm}=\frac{1\pm\cos2\tau}{2}
\ .  \label{prvp0}
\end{equation}
Using these formulas and differentiating with respect to $\tau$,
we further obtain
\begin{align}
\frac{\partial}{\partial\tau}{\>}F_{\alpha}(\tau,0)&=
\frac{\alpha}{1-\alpha}\left(p_{+}^{\alpha-1}-p_{-}^{\alpha-1}\right)\cos2\tau
-\frac{\alpha}{1-\alpha}\left(r_{+}^{\alpha-1}-r_{-}^{\alpha-1}\right)\sin2\tau
\nonumber\\
&=\alpha2^{1-\alpha}uv\bigl[f_{\alpha}(u)-f_{\alpha}(v)\bigr]
\ , \label{prvp01}
\end{align}
where the variables $u=\cos2\tau$ and $v=\sin2\tau$. Since $u>v$
for $\tau\in(0;\pi/8)$ and $u<v$ for $\tau\in(\pi/8;\pi/4)$, the
derivative (\ref{prvp01}) is strictly positive in the former
interval and strictly negative in the latter one. So, the minimal
value of $F_{\alpha}(\tau,0)$ is reached at the end points of the
interval $\tau\in[0;\pi/4]$. In both the points, the function
(\ref{ftpd}) is equal to the right-hand side of (\ref{hal0}). This
bound holds for all pure states and remains valid for mixed states
due to concavity of the entropy (\ref{tslent}).

Let us proceed to conditions for equality. In the domain
(\ref{dotv}), the function $F_{\alpha}(\tau,\varphi)$ takes its
minimum $2\ln_{\alpha}(2)$ only at the points $\tau=\varphi=0$ and
$\tau=\pi/4$, $\varphi=0$. In both the points, one of the
distributions $\{p_{\pm}\}$, $\{q_{\pm}\}$, $\{r_{\pm}\}$ is
deterministic and other two are herewith equiprobable. This is the
only case, when the minimum of $F_{\alpha}(\tau,\varphi)$ takes
place. As it is seen from (\ref{prvp0}), the distribution
$\{q_{\pm}\}$ is inevitably equiprobable for the above two points.
The total domain
$\bigl\{(\tau,\varphi):{\>}\tau\in[0;\pi/2],{\>}\varphi\in[0;2\pi)\bigr\}$
for the state (\ref{pswr}) contains also points, in which the
distribution $\{q_{\pm}\}$ is deterministic. In any case, it is
necessary for reaching the minimum that one of the distributions
be deterministic. In other words, the state $|\psi\rangle$ should
be an eigenstate of one of the observables $\bsg_{x}$, $\bsg_{y}$,
$\bsg_{z}$. Of course, this condition is sufficient as well. We
shall now prove that the inequality (\ref{hal0}) cannot be
saturated with impure states. Let the spectral decomposition of
impure $\bro$ be written as
\begin{equation}
\bro=\lambda_{+}|\psi_{+}\rangle\langle\psi_{+}|+\lambda_{-}|\psi_{-}\rangle\langle\psi_{-}|
\ , \label{spdc}
\end{equation}
where eigenstates are mutually orthogonal and strictly positive
eigenvalues obey the condition $\lambda_{+}+\lambda_{-}=1$. By
concavity of the entropy (\ref{tslent}), we write
\begin{equation}
\sum_{\nu=x,y,z}H_{\alpha}\bigl(\bsg_{\nu}|\bro\bigr)\geq
\lambda_{+}\sum_{\nu=x,y,z}H_{\alpha}\bigl(\bsg_{\nu}|\psi_{+}\bigr)+
\lambda_{-}\sum_{\nu=x,y,z}H_{\alpha}\bigl(\bsg_{\nu}|\psi_{-}\bigr)
\ . \label{nxyz}
\end{equation}
If the sum of $\alpha$-entropies of the state $|\psi_{+}\rangle$
or $|\psi_{-}\rangle$ does not reach the lower bound
$2\ln_{\alpha}(2)$, the left-hand side of (\ref{nxyz}) does not
reach this bound as well. So, the question is reduced to the case,
when the matrix $\bro$ is diagonal with respect to eigenbasis of
either of the $\bsg_{x}$, $\bsg_{y}$, $\bsg_{z}$. For
definiteness, we assume that the $\bro$ commutes with $\bsg_{x}$
and $|\psi_{\pm}\rangle=|x_{\pm}\rangle$. Measuring any of the
$\bsg_{y}$ and $\bsg_{z}$ in the state $\bro$ results in the
equiprobable distribution, whence
$H_{\alpha}\bigl(\bsg_{y}|\bro\bigr)=H_{\alpha}\bigl(\bsg_{z}|\bro\bigr)=\ln_{\alpha}(2)$.
Measuring the $\bsg_{x}$ in the state $\bro$, we obtain outcomes
$\pm1$ with probabilities $\lambda_{\pm}$, respectively. Except
for the two cases, when $\lambda_{+}=1$ or $\lambda_{-}=1$, this
probability distribution is not deterministic and
$H_{\alpha}\bigl(\bsg_{x}|\bro\bigr)>0$. The latter implies that
the sum of three entropies is strictly larger than
$2\ln_{\alpha}(2)$.

Let us consider the standard case $\alpha=1$. Taking the limit
$\alpha\to1^{-}$ in the inequality (\ref{hal0}), we obtain the
known lower bound $2\ln2$ on the standard entropic sum. In this
way, however, conditions for equality are still not resolved.
Nevertheless, we could repeat the above reasons with the function
\begin{equation}
F_{1}(\tau,\varphi)=\sum_{j=\pm}\left(-p_{j}\ln{p}_{j}-q_{j}\ln{q}_{j}-r_{j}\ln{r}_{j}\right)
\ . \label{ft1pd}
\end{equation}
A sketch of the derivation is given below. Similarly to
(\ref{drph}) and (\ref{gadf}), we obtain the formulas
\begin{align}
&\frac{\partial}{\partial\varphi}{\>}F_{1}(\tau,\varphi)=\frac{uv}{2}{\,}\bigl[f_{1}(u)-f_{1}(v)\bigr]
\ , \label{dr1ph0}\\
&f_{1}(u)=\frac{1}{u}{\>}\ln\!\left(\frac{1+u}{1-u}\right)=\sum_{k=0}^{\infty} \frac{2}{2k+1}{\>}u^{2k}
\ ,  \label{dr1ph}
\end{align}
where $u=\sin2\tau\cos\varphi$, $v=\sin2\tau\sin\varphi$. The
function $f_{1}(u)$ monotonically increases, whence
$\partial{F}_{1}(\tau,\varphi)/\partial\varphi>0$ for $0<v<u<1$.
Hence, the minimal value of $F_{1}(\tau,\varphi)$ in the domain
(\ref{dotv}) is reached on the line $\varphi=0$. Differentiating
$F_{1}(\tau,0)$ with respect to $\tau$, we also obtain the formula
\begin{equation}
\frac{\partial}{\partial\tau}{\>}F_{1}(\tau,0)=
uv\bigl[f_{1}(u)-f_{1}(v)\bigr]
\ , \label{prvp11}
\end{equation}
in which $u=\cos2\tau$ and $v=\sin2\tau$. To sum up, we see that
the function $F_{1}(\tau,\varphi)$ takes its minimum $2\ln2$ only
at the points $\tau=\varphi=0$ and $\tau=\pi/4$, $\varphi=0$. As
above, this leads to the claimed conditions for equality.
$\blacksquare$

Theorem \ref{inta01} provides a lower bound on the sum of three
entropies for all $\alpha\in(0;1]$. This bound is tight in the
sense that it is certainly reached with an eigenstate of one of
the complementary observables. Previously, the standard case
$\alpha=1$ has been studied in \cite{jsan93}. In this regard, we
have extended the uncertainty relation for three spin-$1/2$
observables to an entire family of $\alpha$-entropic relations for
all $\alpha\in(0;1]$. In general, a utility of entropic bounds
with a parametric dependence was noted in \cite{maass}. For
example, this dependence allows to find more exactly the domain of
acceptable values for unknown probabilities with respect to known
ones. For the standard case $\alpha=1$, the writers of
\cite{clz11} have derived a stronger bound of state-dependent
form. In their lower bound, the term $2\ln{2}$ is added by the von
Neumann entropy of the $\bro$. It would be of interest to examine
this issue with respect to Tsallis-entropy relations.

\section{Lower bounds on the sum of entropies of degree $\alpha\in(1;+\infty)$}\label{sec4}

In this section, we obtain two connected results. The first result
provides tight lower bounds on the entropic sum for integer
$\alpha\geq2$. The second result presents approximate lower bounds
for non-integer $\alpha\in(1;+\infty)$. To derive the claims, we
will use a lemma. Assuming the expressions
(\ref{xpr})--(\ref{zpr}), we introduce the function
\begin{equation}
G_{\alpha}(\tau,\varphi):=3-\Phi_{\alpha}(p)-\Phi_{\alpha}(q)-\Phi_{\alpha}(r)=
3-\sum_{j=\pm}\left(p_{j}^{\alpha}+q_{j}^{\alpha}+r_{j}^{\alpha}\right)
\ . \label{gdef}
\end{equation}
which is closely related to (\ref{ftpd}). In the present section,
the function $G_{\alpha}(\tau,\varphi)$ will be more convenient.
With the function (\ref{gdef}), we have the following statement.

\newtheorem{lem41}[thm31]{Lemma}
\begin{lem41}\label{intal1}
For integer $\alpha\geq1$, the minimal value of the function
$G_{\alpha}(\tau,\varphi)$ in the domain (\ref{dotv}) is equal to
\begin{equation}
\underset{D}{\min}{\>}G_{\alpha}(\tau,\varphi)=2\left(1-2^{1-\alpha}\right)
{\>.} \label{gmin}
\end{equation}
For integer $\alpha\geq4$, this minimum is reached only at the
points $\tau=\varphi=0$ and $\tau=\pi/4$, $\varphi=0$.
\end{lem41}

{\bf{Proof.}} By doing some simple algebra, we obtain
$G_{1}(\tau,\varphi)\equiv0$, $G_{2}(\tau,\varphi)\equiv1$, and
$G_{3}(\tau,\varphi)\equiv3/2$. These values concur with the
formula (\ref{gmin}). So, we should prove (\ref{gmin}) only for
integer $\alpha\geq4$. Differentiating with respect to $\varphi$,
we obtain
\begin{align}
\frac{\partial}{\partial\varphi}{\>}G_{\alpha}(\tau,\varphi)&=
\frac{\alpha}{2}\left(p_{+}^{\alpha-1}-p_{-}^{\alpha-1}\right)\sin2\tau\sin\varphi-
\frac{\alpha}{2}\left(q_{+}^{\alpha-1}-q_{-}^{\alpha-1}\right)\sin2\tau\cos\varphi
\nonumber\\
&=\alpha2^{-\alpha}uv\bigl[g_{\alpha}(u)-g_{\alpha}(v)\bigr]
{\>}. \label{grph}
\end{align}
For brevity, the result is expressed in terms of the variables
$u=\sin2\tau\cos\varphi$, $v=\sin2\tau\sin\varphi$, and the
function
\begin{equation}
g_{\alpha}(u)=\frac{(1+u)^{\alpha-1}-(1-u)^{\alpha-1}}{u}
\ . \label{gald}
\end{equation}
Except for the boundary lines of the rectangle (\ref{dotv}), we
have $0<v<u<1$. We now claim that the function $g_{\alpha}(u)$
monotonically increases with $u\in(0;1)$ for all integer
$\alpha\geq4$. This fact easily follows from its representation as
a polynomial. Using the binomial formula, it is written as
\begin{equation}
g_{\alpha}(u)=\sum_{k=0}^{\lfloor{\alpha/2}\rfloor-1} 2\binom{\alpha-1}{2k+1} u^{2k}
\ . \label{bi41}
\end{equation}
By $\lfloor{u}\rfloor$, we will mean the floor of real number $u$.
Note that $g_{1}(u)\equiv0$, $g_{2}(u)\equiv2$, and
$g_{3}(u)\equiv4$, whence the derivative (\ref{grph}) vanishes for
$\alpha=1,2,3$. For $\alpha\geq4$, the sum (\ref{bi41}) is not
constant, as non-zero powers are inserted. Further, the
coefficients in (\ref{bi41}) are all strictly positive, whence we
see a monotone increase.

Since the function (\ref{gald}) monotonically increases with
$u\in(0;1)$, the inequality $v<u$ implies
$g_{\alpha}(v)<g_{\alpha}(u)$. In the interior of the rectangle
(\ref{dotv}), therefore, the derivative (\ref{grph}) is strictly
positive. Here, the function $G_{\alpha}(\tau,\varphi)$ increases
with $\varphi$. On the boundary lines $\tau=0$ and $\tau=\pi/4$,
we have $u=v$ and zero derivative (\ref{grph}). These two points
imply that the minimal and maximal values of
$G_{\alpha}(\tau,\varphi)$ in the domain (\ref{dotv}) are reached
on the boundary lines $\varphi=0$ and $\varphi=\pi/4$,
respectively. To find the minimum, we take $\varphi=0$ and write
the expressions (\ref{prvp0}). Differentiating with respect to
$\tau$, we further obtain
\begin{align}
\frac{\partial}{\partial\tau}{\>}G_{\alpha}(\tau,0)&=
-\alpha\left(p_{+}^{\alpha-1}-p_{-}^{\alpha-1}\right)\cos2\tau+
\alpha\left(r_{+}^{\alpha-1}-r_{-}^{\alpha-1}\right)\sin2\tau
\nonumber\\
&=\alpha2^{1-\alpha}uv\bigl[g_{\alpha}(u)-g_{\alpha}(v)\bigr]
\ , \label{prvp1}
\end{align}
where the variables $u=\cos2\tau$ and $v=\sin2\tau$. Of course,
the term $g_{\alpha}(u)$ is constant and the derivative
(\ref{prvp1}) is zero for $\alpha=1,2,3$. For integer
$\alpha\geq4$, the $g_{\alpha}(u)$ monotonically increases, whence
we see the following. As $u>v$ for $\tau\in(0;\pi/8)$ and $u<v$
for $\tau\in(\pi/8;\pi/4)$, the derivative (\ref{prvp1}) is
strictly positive in the former interval and strictly negative in
the latter one. So, the minimal value of $G_{\alpha}(\tau,0)$ is
reached at the end points of the interval $\tau\in[0;\pi/4]$. In
both the points, the function (\ref{gdef}) is equal to the
right-hand side of (\ref{gmin}). $\blacksquare$

Using the statement of Lemma \ref{intal1}, we can obtain a lower
bound on the entropic sum for integer $\alpha\geq2$. The result is
posed in the following way.

\newtheorem{thm41}[thm31]{Theorem}
\begin{thm41}\label{inta12}
Let qubit state be described by density matrix $\bro$. For all
integer $\alpha\geq2$, the entropic sum satisfies
\begin{equation}
H_{\alpha}\bigl(\bsg_{x}|\bro\bigr)+H_{\alpha}\bigl(\bsg_{y}|\bro\bigr)+
H_{\alpha}\bigl(\bsg_{z}|\bro\bigr)\geq2\ln_{\alpha}(2)
\ . \label{hal1}
\end{equation}
For integer $\alpha\geq4$, equality takes place if and only if the
qubit state is an eigenstate of either of the $\bsg_{x}$,
$\bsg_{y}$, $\bsg_{z}$.
\end{thm41}

{\bf{Proof.}} Dividing the right-hand side of (\ref{gmin}) by
$(\alpha-1)$, we get the right-hand side of (\ref{hal1}). This
bound holds for all pure states and remains valid for mixed states
due to concavity of the entropy (\ref{tslent}). For $\alpha=2,3$,
the inequality (\ref{hal1}) is actually saturated with any pure
state. The claim follows from the above equalities
$G_{2}(\tau,\varphi)\equiv1$ and $G_{3}(\tau,\varphi)\equiv3/2$.
Remaining task is to prove conditions for equality in the case of
integer $\alpha\geq4$. In general, this task can be resolved
similarly to conditions for equality in the relation (\ref{hal0})
(see the second part of the proof of Theorem \ref{inta01}). First,
we notice that the $G_{\alpha}(\tau,\varphi)$ takes its minimum
(\ref{gmin}), if and only if one of the distributions
$\{p_{\pm}\}$, $\{q_{\pm}\}$, $\{r_{\pm}\}$ is deterministic.
Hence, the state $|\psi\rangle$ should be an eigenstate of one of
the observables $\bsg_{x}$, $\bsg_{y}$, $\bsg_{z}$. Second, we
prove that the inequality (\ref{hal1}) cannot be saturated with
impure states. We refrain from presenting the details here.
$\blacksquare$

As was mentioned above, for $\alpha=2,3$ the inequality
(\ref{hal1}) is saturated with each pure state. On the other hand,
with an impure state the entropic sum can be increased up to the
maximum $3\ln_{\alpha}(2)$. This upper bound is explained in the
next section. The lower bound of Theorem \ref{inta12} is tight in
the sense that it is always reached with an eigenstate of one of
the complementary observables. It holds for all integer
$\alpha\geq2$. Basing on the result (\ref{gmin}), we can also
obtain an approximate lower bound on the entropic sum for
arbitrary $\alpha\in(1;+\infty)$. Here, functional properties of
$\Phi_{\alpha}(p)$ with respect to the parameter $\alpha$ are
significant. Namely, the quantity $\Phi_{\alpha}(p)$ is a convex
function of $\alpha>0$. Calculating the second derivative, one
actually gives
\begin{equation}
\frac{\partial^{2}\Phi_{\alpha}(p)}{\partial\alpha^{2}}=\sum\nolimits_{p_{j}\neq0}
p_{j}^{\alpha}\bigl(\ln{p}_{j}\bigr)^{2}\geq0
\ . \label{cncph}
\end{equation}
For fixed probabilities, therefore, the quantity (\ref{gdef}) is a
concave function of $\alpha$. Using this concavity, we formulate
the following bound.

\newtheorem{thm42}[thm31]{Theorem}
\begin{thm42}\label{int2}
For all real $\alpha\in(1;+\infty)$ and arbitrary qubit density
matrix $\bro$, the entropic sum satisfies
\begin{equation}
H_{\alpha}\bigl(\bsg_{x}|\bro\bigr)+H_{\alpha}\bigl(\bsg_{y}|\bro\bigr)+
H_{\alpha}\bigl(\bsg_{z}|\bro\bigr)\geq
2{\>}\frac{1-2^{1-\lfloor\alpha\rfloor}}{\alpha-1}
+2^{1-\lfloor\alpha\rfloor}{\,}\frac{\alpha-\lfloor\alpha\rfloor}{\alpha-1}
\ . \label{hal2}
\end{equation}
\end{thm42}

{\bf{Proof.}} Suppose that $\alpha\in(n;n+1)$, where integer
$n=\lfloor\alpha\rfloor\geq1$. The principal point is that the
term $\Phi_{\alpha}(p)$ is a convex function of the parameter
$\alpha$. Therefore, for arbitrary distributions $\{p_{\pm}\}$,
$\{q_{\pm}\}$, $\{r_{\pm}\}$ the term (\ref{gdef}) is concave with
respect to $\alpha$. It follows from the concavity that
\begin{equation}
G_{\alpha}-G_{n}-\left(G_{n+1}-G_{n}\right)(\alpha-n)\geq0
\ , \label{concg}
\end{equation}
for all $\alpha\in[n;n+1]$. Indeed, the left-hand side of
(\ref{concg}) is a concave function of $\alpha$ and vanishes in
both the end points $\alpha=n$ and $\alpha=n+1$. Combining the
inequality (\ref{concg}) with the result (\ref{gmin}), we get
\begin{equation}
G_{\alpha}(\tau,\varphi)\geq
(n+1-\alpha){\,}G_{n}(\tau,\varphi)+(\alpha-n){\,}G_{n+1}(\tau,\varphi)
\geq2\left(1-2^{1-n}\right)+2^{1-n}(\alpha-n)
{\,}. \label{concg1}
\end{equation}
Dividing (\ref{concg1}) by $(\alpha-1)$ completes the proof.
$\blacksquare$

The statement of Theorem \ref{int2} provides a non-trivial lower
bound on the sum of three $\alpha$-entropies. By construction,
this bound is not exact in general. In the limit $\alpha\to1^{+}$,
the right-hand side of (\ref{hal2}) gives $1$, whereas the tight
bound is $2\ln2\approx1.386$. Nevertheless, the lower bound
(\ref{hal2}) is tight for all integer $\alpha\geq2$. In
combination, the relations (\ref{hal0}) and (\ref{hal2}) give
lower bounds on the entropic sum for all real $\alpha>0$. So we
have obtained uncertainty relations in terms of Tsallis'
$\alpha$-entropies for arbitrary positive values of the parameter.
In the case of several observables, entropic bounds are often
given with averaging over the individual entropies
\cite{ww10,ngbw12}. It is also convenient to relate each
$\alpha$-entropy with the value $\ln_{\alpha}(2)$, which
represents the maximum. For all real $\alpha\in(0;1]$ and integer
$\alpha\geq2$, we obtain the tight lower bound on the rescaled
average $\alpha$-entropy
\begin{equation}
\frac{1}{3\ln_{\alpha}(2)}\sum_{\nu=x,y,z}H_{\alpha}\bigl(\bsg_{\nu}|\bro\bigr)\geq\frac{2}{3}
\ . \label{avlb}
\end{equation}
In the left-hand side, the denominator involves $3$ due to
averaging over the three observables and $\ln_{\alpha}(2)$ as a
natural entropic scale. In the range of its validity, the lower
bound (\ref{avlb}) does not depend on the parameter $\alpha$. In a
similar manner, the relation (\ref{hal2}) leads to the
average-entropy lower bound, which is dependent on non-integer
$\alpha>1$.

\section{Some tight upper bounds on the entropic sum in the case of pure states}\label{sec5}

In this section, we study upper bounds on the sum of three
$\alpha$-entropies for complementary qubit observables. In general,
these bounds are essentially depend on a type of considered
states. The completely mixed state is described by density
operator $\bro_{*}=\pen/2$, where $\pen$ denotes the identity
$2\times2$-matrix. Measuring each of the observables $\bsg_{x}$,
$\bsg_{y}$, $\bsg_{z}$ in this state will lead to the equiprobable
distribution. With this distribution, the entropy (\ref{tslent})
takes its maximal value $\ln_{\alpha}(2)$. For all $\alpha>0$ and
arbitrary density matrix $\bro$, we can then write the upper bound
\begin{equation}
\sum_{\nu=x,y,z}H_{\alpha}\bigl(\bsg_{\nu}|\bro\bigr)\leq
\sum_{\nu=x,y,z}H_{\alpha}\bigl(\bsg_{\nu}|\bro_{*}\bigr)
=3\ln_{\alpha}(2)
\ . \label{escm}
\end{equation}
Developing the issue, we ask for upper entropic bounds in the case
of pure states. In line with the method of previous sections, one
will obtain tight bounds from above for real $\alpha\in(0;1]$ and
integer $\alpha\geq2$. Before the derivation, we present some
intuitive reasons that make the result physically reasonable. For
the pure state (\ref{pswr}), the sum of three $\alpha$-entropies
is represented by the function (\ref{ftpd}). Formally, we aim to
maximize this function in the domain (\ref{dotv}). As we have seen
in the proof of Theorem \ref{inta01}, for $\alpha\in(0;1]$ the
maximum is reached on the line $\varphi=\pi/4$. The same remains
valid for integer $\alpha\geq2$, though we have dealt with
(\ref{gdef}) in Lemma \ref{intal1}. Taking $\varphi=\pi/4$ in the
formulas (\ref{xpr}), (\ref{ypr}), and (\ref{zpr}), we obtain the
probabilities
\begin{equation}
p_{\pm}=q_{\pm}=\frac{1\pm{v}}{2}
\ , \qquad
r_{\pm}=\frac{1\pm{u}}{2}
\ ,  \label{prvp2}
\end{equation}
where $u=\cos2\tau$, $v=\sin2\tau/\sqrt{2}$. Obviously, the
variables $u$ and $v$ satisfy the condition
\begin{equation}
u^{2}+2v^{2}=1
\ . \label{uvcd}
\end{equation}
According to (\ref{prvp2}), the distributions $\{p_{\pm}\}$ and
$\{q_{\pm}\}$ should concur for maximizing the entropic sum in the
case of pure states and considered values of $\alpha$. For impure
states, the maximum (\ref{escm}) is reached only if the
probability distributions are all equiprobable and herewith
identical. It is natural to assume that with the terms
(\ref{prvp2}) the maximum takes place, when the distribution
$\{r_{\pm}\}$ also concurs with $\{p_{\pm}\}=\{q_{\pm}\}$, i.e.
$u=v$. Combining the latter with (\ref{uvcd}) gives
$u=v=1/\sqrt{3}$. The $\alpha$-entropy of each of three
probability distributions is then expressed as
\begin{equation}
\wth_{\alpha}=\frac{1}{1-\alpha}{\,}\left\{
\left(\frac{1+1/\sqrt{3}}{2}\right)^{\alpha}+\left(\frac{1-1/\sqrt{3}}{2}\right)^{\alpha}-1
\right\}
{\,}. \label{prone}
\end{equation}
The value $3{\,}\wth_{\alpha}$ gives the sum of three
$\alpha$-entropies. We now claim the following.

\newtheorem{thm51}[thm31]{Theorem}
\begin{thm51}\label{uptal}
Let qubit state be described by ket $|\psi\rangle$. For all real
$\alpha\in(0;1]$ and integer $\alpha\geq2$, the entropic sum obeys
\begin{equation}
H_{\alpha}\bigl(\bsg_{x}|\psi\bigr)+H_{\alpha}\bigl(\bsg_{y}|\psi\bigr)+H_{\alpha}\bigl(\bsg_{z}|\psi\bigr)
\leq3{\,}\wth_{\alpha}
\ . \label{upwh}
\end{equation}
For $\alpha\in(0;1]$ and integer $\alpha\geq4$, equality holds if
and only if the three probability distributions are all, up to
swapping, the pair $\left(1\pm1/\sqrt{3}\right)/2$.
\end{thm51}

{\bf{Proof.}} With the probabilities (\ref{prvp2}), we rewrite the
function (\ref{ftpd}) as
\begin{equation}
\wff_{\alpha}(u,v)=\frac{2^{-\alpha}}{1-\alpha}{\,}
\Bigl\{(1+u)^{\alpha}+(1-u)^{\alpha}+
2(1+v)^{\alpha}+2(1-v)^{\alpha}-3\cdot2^{\alpha}
\Bigr\}
{\>}. \label{ftpd2}
\end{equation}
When $\tau\in[0;\pi/4]$, the variables $u$ and $v$ lie in the
interval $[0;1]$. The function (\ref{ftpd2}) should be maximized
in this interval under the condition (\ref{uvcd}). It follows from
(\ref{uvcd}) that $du/dv=-2v/u$. Differentiating (\ref{ftpd2})
with respect to $v$, we then obtain
\begin{align}
&\frac{\alpha2^{-\alpha}}{1-\alpha}
\left\{-\frac{2v}{u}\left[(1+u)^{\alpha-1}-(1-u)^{\alpha-1}\right]+
2\left[(1+v)^{\alpha-1}-(1-v)^{\alpha-1}\right]
\right\}
\nonumber\\
&=\alpha2^{1-\alpha}v\bigl[f_{\alpha}(u)-f_{\alpha}(v)\bigr]
\label{ftpd21}\\
&=\frac{\alpha2^{1-\alpha}v}{\alpha-1}{\,}\bigl[g_{\alpha}(u)-g_{\alpha}(v)\bigr]
\ . \label{ftpd22}
\end{align}
Here, the functions $f_{\alpha}(u)$ and $g_{\alpha}(u)$ were
defined in (\ref{gadf}) and (\ref{gald}). For $\alpha\in(0;1)$,
the expression (\ref{ftpd21}) is convenient. We have seen above
that the function $f_{\alpha}(u)$ monotonically increases. So, the
derivative is strictly positive for $u>v$ and strictly negative
for $u<v$. The maximum is reached for $u=v$, whence each of the
three entropies takes the value (\ref{prone}). Of course, the
function (\ref{ftpd2}) becomes the right-hand side of
(\ref{upwh}). Taking the limit $\alpha\to1^{-}$ in the inequality
(\ref{upwh}), we obtain the upper bound $3{\,}\wth_{1}$ with
\begin{equation}
\wth_{1}=-\frac{1+1/\sqrt{3}}{2}{\,}\ln\!\left(\frac{1+1/\sqrt{3}}{2}\right)
-\frac{1-1/\sqrt{3}}{2}{\,}\ln\!\left(\frac{1-1/\sqrt{3}}{2}\right)
\ . \label{prone1}
\end{equation}
For proving the conditions for equality, more detailed analysis is
required. In the interval $u,v\in[0;1]$, we should maximize the
function
\begin{equation}
\wff_{1}(u,v)=3\ln2-\frac{1+u}{2}{\,}\ln(1+u)-\frac{1-u}{2}{\,}\ln(1-u)-(1+v)\ln(1+v)-(1-v)\ln(1-v)
{\,}, \label{ftpd11}
\end{equation}
under the condition (\ref{uvcd}). By differentiating with respect
to $v$, we then have the right-hand side of (\ref{ftpd21}) with
$\alpha=1$ in terms of the function $f_{1}(u)$ defined in
(\ref{dr1ph}). The function $f_{1}(u)$ monotonically increases as
well. Therefore, we can repeat all the above reasons including the
condition $u=v$ for reaching the maximum.

For integer $\alpha\geq2$, we will use (\ref{ftpd22}). As
$g_{2}(u)\equiv2$ and $g_{3}(u)\equiv4$, the quantity
(\ref{ftpd22}) is zero. This is a manifestation of the fact that
for $\alpha=2,3$ the left-hand side of (\ref{upwh}) is constant
irrespectively of the ket $|\psi\rangle$. Taking (\ref{uvcd}), we
actually have $\wff_{2}=1$ and $\wff_{3}=3/4$. Since
$\wth_{2}=1/3$ and $\wth_{3}=1/4$, the relation (\ref{upwh}) holds
for $\alpha=2,3$. For integer $\alpha\geq4$, the function
$g_{\alpha}(u)$ monotonically increases. Hence, we again have the
result $u=v$. It implies that the the right-hand side of
(\ref{upwh}) gives the maximum.

We shall now prove conditions for equality. For all real
$\alpha\in(0;1]$ and integer $\alpha\geq4$, the maximum
$3{\,}\wth_{\alpha}$ in the domain (\ref{dotv}) is reached only
when $\varphi=\pi/4$ and also $u=v$, i.e.
$2\tau=\arctan\sqrt{2}\approx0.955$. In this point, we have the
probabilities
\begin{equation}
p_{\pm}=q_{\pm}=r_{\pm}=\frac{1\pm1/\sqrt{3}}{2}
\ . \label{pqr3}
\end{equation}
The formula (\ref{dotv}) gives the domain over which the
optimization has been performed. Right after (\ref{dotv}), we have
described those maps that allow to reduce the total domain
$\bigl\{(\tau,\varphi):{\>}\tau\in[0;\pi/2],{\>}\varphi\in[0;2\pi)\bigr\}$
just to (\ref{dotv}). By inverting these maps, the point
$2\tau=\arctan\sqrt{2}$, $\varphi=\pi/4$ will generate other
points in which the inequality (\ref{upwh}) is saturated. In all
the points, each of the pairs $\{p_{\pm}\}$, $\{q_{\pm}\}$,
$\{r_{\pm}\}$ is, up to swapping, the pair (\ref{pqr3}).
$\blacksquare$

\begin{figure}
\includegraphics{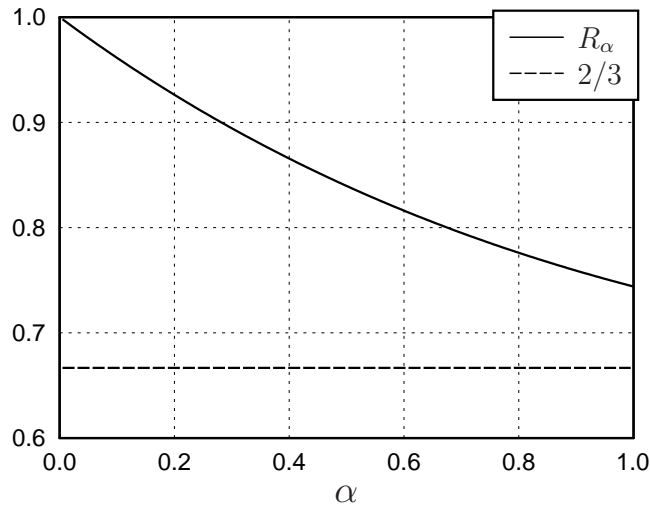}
\caption{The left-hand and right-hand sides of (\ref{loup12}) for real $\alpha\in(0;1]$.}
\label{fig01}
\end{figure}

In the case of pure states, the statement of Theorem \ref{uptal}
provides tight upper bounds on the entropic sum for all real
$\alpha\in(0;1]$ and integer $\alpha\geq2$. Previously, these
bounds have been motivated by some plausible reasons. It seems
that our method cannot be applied to other values of $\alpha$. Let
us take the entropic value, which is both averaged over the
individual ones and rescaled by the denominator $\ln_{\alpha}(2)$.
Combining (\ref{avlb}) and (\ref{upwh}), we obtain the relation
\begin{equation}
\frac{2}{3}\leq
\frac{1}{3\ln_{\alpha}(2)}\sum_{\nu=x,y,z} H_{\alpha}\bigl(\bsg_{\nu}|\psi\bigr)
\leq{R}_{\alpha}=\frac{\wth_{\alpha}}{\ln_{\alpha}(2)}
\ , \label{loup12}
\end{equation}
which is shown for real $\alpha\in(0;1]$ and integer
$\alpha\geq2$. The relative quantity $R_{\alpha}$ gives an upper
bound on the rescaled average $\alpha$-entropy in the case of pure
states. Both the sides of (\ref{loup12}) are tight in the sense
that they are reached under the certain conditions for equality.
Hence, we can describe the band, in which the rescaled average
$\alpha$-entropy ranges. For $\alpha\in(0;1]$, this band is shown
on Fig. \ref{fig01}. The lower bound is constant, whereas the
upper bound monotonically decreases with $\alpha$. So, the band is
reducing with growth of $\alpha$. Although the value $\alpha=0$
itself is not used, we have $R_{\alpha}\to{1}$ in the limit
$\alpha\to0^{+}$. For $\alpha=1$, the upper bound becomes
$\wth_{1}/\ln2\approx0.744$. It can be interpreted as an evidence
for sensitive in quantifying the complementarity. With small
values of the parameter $\alpha$, the average $\alpha$-entropic
measure seems to be more sensitive. For integer $\alpha\geq2$, the
following can be said. As was mentioned above, the right-hand side
of (\ref{loup12}) concurs with the left-hand one for $\alpha=2,3$.
Further, the right-hand side of (\ref{loup12}) increases with
integer $\alpha$. For instance, we calculate $R_{4}\approx0.698$,
$R_{5}\approx0.741$, $R_{6}\approx0.784$, $R_{7}\approx0.823$,
$R_{8}\approx0.857$, $R_{9}\approx0.885$, and
$R_{10}\approx0.909$. For such values of $\alpha$, therefore, the
average $\alpha$-entropic measure is also enough sensitive in a
relative scale. In general, this issue deserves further
investigations.

\section{Conclusion}\label{sec6}

We have obtained new uncertainty and certainty relations for the
Pauli observables in terms of the Tsallis $\alpha$-entropies.
These entropies form an especially important one-parametric
extension of the Shannon entropy. The uncertainty and certainty
relations are respectively expressed as lower and upper bounds on
the sum of three $\alpha$-entropies. Lower bounds on the entropic
sum are given for arbitrary $\alpha>0$. For real $\alpha\in(0;1]$
and integer $\alpha\geq2$, the presented bounds are tight in the
sense that they can certainly be saturated. The conditions for
equality in the relations are obtained as well. For non-integer
$\alpha>1$, we have presented approximate lower bounds on the
entropic sum. Approximate bounds are based on the tight bounds for
integer $\alpha$ and concavity properties with respect to the
parameter $\alpha$. In the case of pure states, tight upper bounds
on the sum have been obtained for all $\alpha\in(0;1]$ and integer
$\alpha\geq2$. These bounds have previously been explained with
some intuitive reasons. Our method seems to be insufficient for
obtaining upper bounds with non-integer $\alpha>1$. In principle,
this issue could be studied by direct numerical calculations.
Indeed, nonlinear optimization problems with no immediate solution
are often arisen in maximizing information-theoretical quantities
\cite{fperes96}. We also note that our results were all tested
numerically. For any number of mutually unbiased bases in finite
dimensions, lower bounds on the sum of Shannon entropies have been
derived in \cite{molm09}. Bounds of such a kind could be obtained
with use of Tsallis' entropies. This issue will be considered in a
following work.

\acknowledgments
The present author is grateful to anonymous referees for useful comments.

\end{document}